\documentclass{article}

\usepackage{url}
\usepackage{amssymb}

\usepackage{epsfig}
\usepackage{wrapfig}
\usepackage{subfigure}
\usepackage{rotating}
\usepackage{moreverb}
\usepackage{fancyvrb}

\def\bd{\begin{description}}
\def\ed{\end{description}}
\def\bc{\begin{center}}
\def\ec{\end{center}}
\def\bq{\begin{quote}}
\def\eq{\end{quote}}
\def\bi{\begin{itemize}}
\def\ei{\end{itemize}}
\def\be{\begin{enumerate}}
\def\ee{\end{enumerate}}
\def\ba{\begin{array}}
\def\ea{\end{array}}

\newcommand{\chr}{{CHR}}
\newcommand{\true}{{\it true}}

\newcommand{\false}{{\it false}}
\newcommand{\simp}{{\; \Leftrightarrow \;}}

\newcommand{\prop}{{\; \Rightarrow\; }}
\newcommand{\com}{\rule{1pt}{7pt}}

\newtheorem{theorem}{\bf Theorem}
\newtheorem{lemma}{\bf Lemma}
\newtheorem{example}{\bf Example}%[section]
\newtheorem{definition}{\bf Definition}%[section]

\newcommand{\qed}{\nobreak \ifvmode \relax \else
     \ifdim\lastskip<1.5em \hskip-\lastskip
      \hskip1.5em plus0em minus0.5em \fi \nobreak
\linethickness{.5pt}\framebox(4,4){}\fi}

\begin{document}

\date{}

\title{A Devil's Advocate against Termination of Direct Recursion}

\author{{Thom Fr{\"u}hwirth}\\
           {University of Ulm, Germany}\\
           {www.informatik.uni-ulm.de/pm/mitarbeiter/fruehwirth/}}

\maketitle

\begin{abstract}
A devil's advocate is one who argues against a claim, 
not as a committed opponent but in order to determine the validity of the claim.
We are interested in a devil's advocate that argues against termination of a program. He does so by producing a maleficent program that can cause the non-termination of the original program. 
By inspecting and running the malicious program, one may gain insight into the potential reasons for non-termination and produce counterexamples for termination. 

We introduce our method using the concurrent programming language Constraint Handling Rules (CHR). 
Like in other declarative languages, non-termination occurs through unbounded recursion.
Given a self-recursive rule, we automatically generate one or more devil's rules from it.
The construction of the devil's rules is straightforward and involves no guessing. 
The devil's rules can be simple. For example, they are non-recursive for rules with single recursion. 

We show that the devil's rules are maximally vicious in the following sense:
For any program that contains the self-recursive rule and for any infinite computation through 
that rule in that program,
there is a corresponding infinite computation with the recursive rule and the devil's rules alone.
In that case, the malicious rules serve as a finite witness for non-termination.
On the other hand, if the devil's rules do not exhibit an infinite computation, the recursive rule is unconditionally terminating.
We also identify cases where the static analysis of the devil's rule decides termination or non-termination of the recursive rule.
\end{abstract}

The final version of this paper is \cite{fruhwirth2015devil}.

\section{Introduction}

It is well known that termination is undecidable for Turing-complete programming languages. There is a long tradition in research on analysis methods to tame the problem by semi-automatic or approximative approaches.
Here we turn the problem around - we look at the dual problem of non-termination. We present a devil's advocate algorithm that argues against termination. 
It produces a maleficent program that can cause the non-termination of the original program. 
The devil's program may be simpler than the orginal one, for example, it can be non-recursive and thus terminating.
The malicious program can form an alternative basis for dynamic and static termination analysis.
When it is run, it can be quite useful in debugging, providing counterexamples for termination. 
We can also derive conditions for both termination and non-termination, as we will show in this paper.

We introduce our devil's advocate method in the programming language Constraint Handling Rules (CHR). 
CHR is a practical concurrent, declarative constraint-based language and versatile computational formalism at the same time.
This allows us to operate on a high level of abstraction, namely first-order predicate logic. 
In this paper, we only consider direct recursion. We think that this setting is best for introducing and demonstrating
our novel approach to (non-)termination analysis. At the same time, we are confident that our devil's advocate technique carries over to other types of recursion and in general to traditional programming languages with while-loops as well.

The following program for determining if a number is even 
will serve as a running example in our paper. 
It also serves as a first overview of the characteristic features of the CHR language.
\begin{example}\label{even1}
{\rm 
In CHR, we use a first-order logic syntax, but
variable names start with upper-case letters, while function and predicate
symbols start with lower-case letters.
Predicates will be called constraints.
The built-in binary infix constraint symbol {\em =} stands for syntactical equality.
For the sake of this example, numbers are expressed in successor notation.
The user-defined unary constraint {\em even} can be implemented by two rules

\medskip
$\begin{array}{l}

\mathit{even(X)} \simp X{=}0 \; \com \; \mathit{true}.\\
\mathit{even(X)} \simp X{=}s(Y) \; \com \; Y{=}s(Z) \land \mathit{even(Z)}.

\end{array}$
\medskip

\noindent The first rule says that $X$ is even if it equals the number $0$.
The recursive rule says that $X$ is even, if it is the
successor of some number $Y$, and then the predecessor of this number
$Z$ is even. $X{=}s(Y)$ is a guard, a precondition for the applicability of the rule.
It serves as a test, while $Y{=}s(Z)$ in the right-hand-side of the rule asserts an equality.

In logical languages like CHR, variables cannot be overwritten, but they can be without value (unbound). For example, if $X$ is $s(s(A))$, then it will satisfy the guard, and $Y$ will be $s(A)$. If $X$ is unbound, then the guard does not hold. If the variable $X$ later becomes (partially) bound in a syntactic equality, the
computation of {\em even} may resume.

CHR is a committed-choice language, i.e. there is no backtracking in the rule applications. 
Computations in CHR are sequences of rule applications starting with a query and ending in an answer.
To the query {\em even(0)} the first rule applies, the answer is {\em true}.
The query {\em even(N)} delays, since no rule is applicable. The answer is the query itself.
To the query {\em even(s(N))} the recursive rule is applicable once, the (conditional) answer is 
$N{=}s(N') \land \mathit{even(N')}$. 

For the recursive rule of {\em even}, the devil's advocate just constructs the non-recursive propagation rule
\[\mathit{even(X)} \prop X{=}s(Y).\]
(In general it may not be that simple.)
In a propagation rule, the left-hand-side constraint is not removed when the rule is applied. So this devil's rule observes occurrences of {\em even(X)} and maliciously adds $X{=}s(Y)$. This will trigger another application of the recursive rule (or lead to an inconsistency if $X$ is $0$). Thus there exist programs in which the recursive rule does not terminate. 

Moreover, as we will show, any non-termination of {\em even} in any program
is in essence characterized by the behavior of its devil's rule.
Conversely, the devil's rule rather bluntly tells us that 
termination is ensured if eventually there is a recursive goal $\mathit{even(X)}$, 
where the variable $X$ is different from $s(Y)$, 
including the case where $X$ is unbound. 
 
In this minimalistic example, the recursive rule alone suffices to produce non-termination.
The query
$\mathit{even(N)} \land \mathit{even(s(N))}$ will not terminate.
Applying the recursive rule to $\mathit{even(s(N))}$ leads to 
$\mathit{even(N)} \land N{=}s(N') \land \mathit{even(N')}$.
Since $N{=}s(N')$, the rule can now be applied to $\mathit{even(N)}$ and so on ad infinitum.
} %rm
\end{example}

\medskip
\noindent{\bf Related Work.}
While there is a vast literature on proving termination 
(one may start with \cite{podelski} and with \cite{pilozzi} for CHR),
proving non-termination has only recently come to the attention of program analysis research. 

Non-termination research can be found for term rewriting systems \cite{giesl05,payet08},
logic programming languages like Prolog \cite{payet06,deschreye} and XSB \cite{kifer}, % ,shen03
constraint logic programming languages (CLP) \cite{mesnardcp}
and imperative languages like Java \cite{popl2008,Brockschmidt,Velroyen,mesnardjava}. % and C.

In \cite{deschreye} non-termination of Prolog is proven by statically checking for loops in a finite abstract computation tree derived from moded queries. Similarily, \cite{kifer} studies the problem of non-termin\-ation in tabled logic engines with subgoal abstraction, such as XSB. The algorithms proposed analyse forest logging traces and output sequences of tabled subgoal calls that are the likely causes of non-terminating cycles.

The papers \cite{mesnardcp, mesnardjava} give a criterion for detecting 
non-terminating atomic queries with respect to binary CLP rules.
The approach is based on abstracting states by so-called filters and proving a recurrence. 
The recurrence criterion is similar to the one in \cite{popl2008}.

For Java, the approach is often to translate into declarative languages and formalisms, e.g.
into term rewriting systems \cite{Brockschmidt},
into logical formulae \cite{Velroyen}
and into CLP \cite{mesnardjava}.
However, these translations are abstractions 
and in general this results in a loss of accuracy.
The method in \cite{popl2008} for imperative languages such as Java is incomplete because the loop must be periodic.
So it cannot deal with nested loops. 

Overall, most research on non-termination can be seen as being based on the approach
that is explained in \cite{popl2008}. 
It is a combination of dynamic and static analysis.
One searches for counterexamples for termination. First, one dynamically enumerates a certain class of candidate execution paths (computations) until a  state is re-entered. This has the drawback of combinatorial explosion in the number of paths. Then the candidate paths are checked if they contain a loop, i.e. a syntactic cycle. 

The check amounts to proving the existence of a so-called recurrence set of states 
(transition invariants on states that are visited infinitely often). 
This problem is formulated as a constraint satisfaction problem
and it is equivalent to the one for invariant generation.

As a reviewer of this paper has pointed out, 
the recent paper \cite{chen2014proving} also builds on recurrence sets, but avoids the need for periodicity in the non-terminating computation. This work is done in the context of a simple imperative while-loop language and the utilization of tools for proving safety properties. The approach performs an underapproximation search of the program to synthesize a reachable non-terminating loop, i.e. to produce an abstraction of the program with assumptions that lead to non-termination.  The algorithms are quite different from our straightforward construction, but the underlying insights into the problem of non-termination seem similar. In particular, this approach, like ours, also produces a (kind of abstracted) program as a witness for non-termination.
A further investigation of the relationship between the methods seems warranted.

\medskip
\noindent {\bf Our Devil's Advocate Method.}
To the best of our knowledge our technique of constructing a malicious program by a devil's advocate algorithm
is novel. 
The construction of the devil's rules is a straightforward program transformation. 
Our approach works well in a concurrent setting.
Our proposed methodology is fully static, it avoids the exploration of all possible paths in a program's execution. 
Furthermore, there is no need for guessing transition invariants, because they are readily encoded in the malicious program.
Our approach covers both periodic, aperiodic and nested loops (in the form of direct recursion), because there is no need to detect syntactic cycles. 

The constructed devil's rules immediately give rise to a non-terminating computation, if there exists one at all.
This maximally vicious computation is the abstraction of all non-terminating computations, 
but at the same time it is an executable program. This is made possible by the use of constraints as abstraction mechanism.

Our approach differs from most existing ones in that we explore non-terminating executions that are essential in that they are independent of the program context. The devil's rules exactly characterize what such a context in essence has to do in order to cause non-termination. Such a maximally vicious program is more general and more concise than characterizing the typically infinite set of queries that would lead to (non-)termination. The the execution of the devil's rules may even correspond to query of infinite size while having a compact finite representation.

The exemplary criterions for (non-)termination that we will
derive for the devil's rules are similar to those that can be found in the literature.
In principle, a transition invariant is defined that is a sufficient condition for the property at hand.
However, in contrast to other research, we do not have to search for invariants or guess them by a heuristic or a process of abstraction steps. Our invariants can be readily derived from the devil's rules, 
that in turn are built from the culprit recursive rules.

\medskip
\noindent{\bf Outline of the Paper.}
In the Preliminaries we introduce syntax and semantics of Constraint Handling Rules (CHR).
In Section 3, we define the construction of malicious rules from directly recursive simplification rules.
They give rise to maximally vicious computations, that never terminate successfully. 
We show that each non-terminating computation of the recursive rule contains a vicious computation.
In Section 4, we look at static analysis of the devil's rules. We propose sufficient conditions for termination and non-termination of vicious computations.
In Section 5, we address the simplification of devil's rules and give some more extended examples for our devil's advocate approach,
before we conclude the paper.
Readers who want a quick overview of the devil's advocate method can skip the proofs.

\section{Preliminaries}\label{sec:chr}

In this section we give an overview of syntax and semantics for Constraint
Handling Rules (CHR)~\cite{chrbook}.
We assume basic familiarity with first-order predicate logic and state transition systems.

\medskip {\bf Abstract Syntax of \chr.}
Constraints are distinguished predicates of first-order predicate logic.
We use two disjoint sets of predicate symbols (or: constraint names) for two different
kinds of constraints: {\em built-in (or: pre-defined) constraints} 
which are handled by a given constraint solver,
and
{\em user-defined (or: CHR) constraints} 
which are defined by the rules in a CHR program.
A {\em \chr\ program} is a finite set of rules.  
There are three kinds of rules:

\begin{center}
\begin{tabular}{ll}
{\em Simplification rule:} & {\it r} $: \; H \simp C \; \com \; B,$ \\
{\em Propagation rule:} & {\it r} $: \; H \prop C \; \com \; B,$ \\
{\em Simpagation rule:} & {\it r} $: \; H_1 \setminus H_2 \simp C \; \com \; B,$ \\
\end{tabular}
\end{center}

\noindent where {\em r:} is an optional, unique identifier of a rule, the {\em head}  %! drop Name?
denoted by $H$, $H_1$ and $H_2$ is a non-empty conjunction of user-defined constraints, the
{\em guard} $C$ is a conjunction of built-in constraints, and the {\em body} $B$
is a goal. A {\em goal (or: query)} is a conjunction of built-in and CHR
constraints.

Conjuncts can be permuted since conjunction is associative and commutative.
We will, however consider conjunction not to be idempotent, since we allow for duplicates, i.e. multiple occurrences of user-defined constraints. 
The empty conjunction is denoted by the built-in constraint $\true$, which is the neutral element of the conjunction operator $\land$. 
A trivial guard expression ``\texttt{true |}'' can be omitted from a rule.

When it is convenient, we allow for {\em generalized simpagation rules}. In such rules,
either $H_1$ or $H_2$ may be empty. 
If $H_1$ is empty, we may write the rule as a simplification rule. 
If $H_2$ is empty, we may write it as a propagation rule.

\medskip {\bf Abstract Operational Semantics of \chr.}\label{sec:chr:semantics}
The operational semantics of CHR is given by the state transition system in Fig. \ref{trans}.
In the figure, all single upper-case letters except $P$ are meta-variables that stand for goals.
Let $P$ be a CHR program. 
Let the variables in a disjoint variant of a rule be denoted by $\bar x$.
Let $CT$ be a complete and decidable constraint theory for the built-in constraints, %!
including the trivial $\true$ and $\false$ as well as syntactical equivalence $=$.
For a goal $G$, the notation $G_{bi}$ denotes the built-in constraints of $G$
and $G_{ud}$ denotes the user-defined constraints of $G$.

A {\em disjoint (or: fresh) variant} of an expression is obtained by uniformly replacing its
variables by different, new (fresh) variables. 
A {\em variable renaming} is a bijective function over variables.
\begin{figure}%[htb]
\begin{center}
  {\bf Simplify}\\[5pt]
  \begin{tabular}{l@{\quad}l}
    If   & %\parbox[t]{.8\textwidth}{\raggedright
      $(r: \; H \Leftrightarrow C \; \com \; B)$ is a disjoint variant of
      a rule in $P$\\ %!- with variables $\bar x$\\
    and  & $CT \models %! \exists (G_{bi}) \land 
			\forall (G_{bi} \to \exists \bar x (H{=}H' \land C))$\\
    then & $({H'} \land G) \mapsto_r (B \land  G
        \land H{=}H' \land C)$
    %$\begin{array}[c]{cc}
  \end{tabular}

  \medskip

  {\bf Propagate}\\[5pt]
  \begin{tabular}{l@{\quad}l}
    If   & %\parbox[t]{.8\textwidth}{\raggedright
      $(r: \; H \Rightarrow C \; \com \; B)$ is a disjoint variant of
      a rule in $P$\\ %!-  with variables $\bar x$\\
    and  & $CT \models %! \exists (G_{bi}) \land 
			\forall (G_{bi} \to \exists \bar x (H{=}H' \land C))$\\
    then & $(H' \land G) \mapsto_r (H' \land B \land G
        \land H{=}H' \land C)$
    %$\begin{array}[c]{cc}
  \end{tabular}

  \medskip

  {\bf Simpagate}\\[5pt]
  \begin{tabular}{l@{\quad}l}
    If   & %\parbox[t]{.8\textwidth}{\raggedright
      $(r:  H_1 \setminus H_2 \Rightarrow C \; \com \; B)$ is a disjoint variant of
      a rule in $P$\\ % with variables $\bar x$\\
    and  & $CT \models %! \exists (G_{bi}) \land 
			\forall (G_{bi} \to \exists \bar x (H_1{\land}H_2){=}(H'_1{\land}H'_2) \land C))$\\
    then & $(H'_1 \land H'_2 \land G) \mapsto_r$\\
        & $(H'_1 \land B \land G
        \land (H_1{\land}H_2){=}(H'_1{\land}H'_2) \land C)$
    %$\begin{array}[c]{cc}
 \end{tabular}

\caption{Transitions of Constraint Handling Rules}
\label{trans}
\end{center}
\end{figure}

Starting with a given initial state (or: query), 
CHR rules are applied exhaustively, until a fixed-point is reached.
A rule is {\em applicable}, if 
its head constraints are matched by constraints in the current goal
one-by-one and if, under this matching, 
the guard of the rule is logically implied by the built-in constraints in the goal.
An expression of the form 
$CT \models %! \exists (G_{bi}) \land 
			\forall (G_{bi} \to \exists \bar x (H{=}H' \land C))$
is called {\em applicability condition}.
Any one of the applicable rules can
be applied in a transition, and the application cannot be undone, it is committed-choice.

A simplification rule 
$H \simp C \; \com \; B$
that is applied
{\em removes} the user-defined constraints matching $H$ and replaces them by $B$ provided the guard
$C$ holds.  
A propagation rule 
$H \prop C \; \com \; B$
instead {\em keeps} $H$ and {\em adds} $B$.
A simpagation rule
$H_1 \setminus H_2 \simp C \; \com \; B$
keeps $H_1$, removes $H_2$ and adds $B$.
If new constraints arrive, rule applications are restarted.

States are goals.
In a transition (or: computation step) 
$S \mapsto_r T$, $S$ is called {\em source state} and $T$ is called {\em target state}.
A {\em computation} of a goal $G$ in a program $P$
is a connected sequence
$S_i \mapsto S_{i+1}$ beginning with
the initial state $S_0=G$ 
and ending in a final state (or: answer) or the sequence is non-terminating (or: diverging). 
The length of a computation is the number of its computation steps.
The notation ${\mapsto_P}^n$ denotes a finite computation of length $n$ 
where rules from $P$ have been applied.
Given a computation starting with $S_0$ in which
a state $S_i$ with $(0 \leq i)$ occurs, then
the computation up to $S_i$ 
is a {\em prefix} of the computation.

Note that built-in constraints in a computation are accumulated, i.e. added but never removed, while user-defined constraints can be added as well as removed.

In the transitions of the abstract semantics as given,
there are two sources of {\em trivial non-termination}.
For simplicity, we have not made their avoidance explicit in the transitions of the abstract semantics.
(Concurrency is also not made explicit in the semantics given.)

First, if the built-in constraints $G_{bi}$ in a state are inconsistent (or: unsatisfiable), 
any rule could be applied to it, since
the applicability condition trivially holds since the premise of the logical implication is false.
We call such a state {\em failed}.
Non-termination due to failed states is avoided by requiring $G_{bi}$ to be consistent when a rule is applied. 
In other words, any state with inconsistent built-in constraints is a (failed) final state.

Second, a propagation rule could be applied again and again, since it does not remove any constraints and thus its applicability condition always continues to hold after the rule has been applied (due to CHR's monotonicity).
This non-termination is avoided by applying a propagation rule at most once to the same user-defined constraints.
Note that syntactically identical user-defined constraints are not necessarily the same, since we allow for duplicates.
In implementations, each user-defined constraint has a unique identifier, and only constraints with the same identifier are considered to be the same for this purpose.

\section{Devil's Advocate against Termination of Direct Recursion}

Our devil's advocate algorithm constructs one or more malicious rules from a given recursive rule.
The idea behind these devil's rules can be explained as follows:
A devil's rule observes the computation. 
When it sees constraints that could come from the body of the recursive rule, 
it suspects the recursive rule has just been applied.
It then maliciously adds the constraints necessary to trigger another recursive step by making the recursive rule applicable.

We therefore first prove that this interplay between the recursive rule and the devil's rule can only lead to non-termination or a failed state. We call such a computation maximally vicious. This is because the devil's rules capture the essence of any non-termination of the recursive rule, no matter in which program. 
Even if the devil's rule is not present, every infinite computation through the recursive rule will remove and add some constraints in exactly the same way as the devil's rule would do. 

We therefore prove a second claim, namely that any non-terminating computation without the devil's rule contains the maximally vicious computation with the devil's rule.
Therefore, if the devil's rules do not exhibit an infinite computation, the recursive rule is unconditionally terminating.

\subsection{Direct Recursion, Devil's Advocate and Devil's Rules}

In this paper, we consider direct recursion expressed by simplification rules.
From them, our devil's advocate algorithm will construct devil's rules.   

\begin{definition}{\rm
A CHR rule is {\em direct recursive (or: self-recursive)} if the head and the body of the rule have common predicate symbols. 
A constraint is direct recursive if its predicate symbol occurs in the head and in the body of a rule.
}
\end{definition}

An overlap is a conjunction built from two goals, where one or more constraints from different goals are equated pairwise.
\begin{definition}{\rm
Given two conjunctions of constraints $A$ of the form $A_1 \land A_2$ 
 and $B$ of the form $B_1 \land B_2$,
where $A_2$ and $B_2$ are non-empty conjunctions.
An {\em overlap $A \diamond B$ at the common constraints $A_2$ and $B_2$}  %! common, not shared
is a conjunction of the form
$A_1 \land A_2 \land B_1 \land A_2{=}B_2$.  
The goal $A_2 \land A_2{=}B_2$ is called the {\em common part} of the overlap. %!
}
\end{definition}
Note an overlap is only possible if the two goals have common predicate symbols. If there are more than two such constraints, there are several overlaps.

Now the rules constructed by the devil's advocate come into play. 
\begin{definition}\label{devilsrule}
{\rm
Given a self-recursive simplification rule $r$ of the form 
\[r: H \simp C \; \com \; B_{bi} \land B_{ud},\]   %! in head normal form
where $B_{bi}$ denotes the built-in constraints
and $B_{ud}$ denotes the user-defined constraints comprising the rule body $B$.
Let $r'$ be a disjoint variant of the rule $r$ of the form
\[r': H' \simp C' \; \com \; B'_{bi} \land B'_{ud}.\]

For each overlap $(B_{ud} \diamond H')$    
at the common constraints $O_{B_{ud}}$ and $O_{H'}$, we generate a devil's rule.  

A {\em devil's rule $d$ for the rule $r$} is a generalised simpagation rule of the form
\[d: O_{B_{ud}} \setminus R_{B_{ud}} \simp C \land B_{bi} \; \com \; C' \land O_{B_{ud}}{=}O_{H'} \land R_{H'}\]
where 
$B_{ud}{=}(O_{B_{ud}}\land R_{B_{ud}})$ and $H'{=}(O_{H'} \land R_{H'})$.
}
\end{definition}
Note that the effect of the devil's rule is to manipulate exactly those constraints that occur in its recursive rule. In particular, it in effect replaces all user-defined body constraints of the recursion by the head constraints that are needed for the next recursive step.

\begin{example}\label{even2} %! user tabular or array format layout
{\rm 
We continue with Example \ref{even1} and its rule
\[\mathit{even(X)} \simp X{=}s(Y) \; \com \; Y{=}s(Z) \land \mathit{even(Z)}.\]
The rule for {\em even} is direct single recursive, 
so there is only one overlap and the resulting single devil's rule is not recursive.
Also, the goal $R_{B_{ud}}$ is empty, 
thus we can write the generated devil's rule as a propagation rule
\[\mathit{even(Z)}{\Rightarrow}X{=}s(Y) \land Y{=}s(Z) \com {X'}{=}s(Y') \land \mathit{even(Z){=}even(X')}.\]
In Section \ref{examples} we will simplify this rule into $\mathit{even(X)} \prop X{=}s(Y)$.
} %rm
\end{example}
\begin{example}\label{traverse} %! user tabular or array format layout
{\rm 
Consider a rule scheme for tree traversal of the form
\[\mathit{traverse}(node(L,V,R)) \simp  C \; \com \; B \land \mathit{traverse}(L) \land \mathit{traverse}(R).\]
It yields two devil's rules that are variants of each other

\medskip
$\begin{array}{l}
\mathit{traverse}(L) \setminus \mathit{traverse}(R) \simp  C \land B \; \com \;\\
   \indent C' \land (\mathit{traverse}(L){=}\mathit{traverse}(node(L',V',R'))),\\
\mathit{traverse}(R) \setminus \mathit{traverse}(L) \simp  C \land B \; \com \;\\
   \indent C' \land (\mathit{traverse}(R){=}\mathit{traverse}(node(L',V',R'))).\\
\end{array}$
\medskip

} %rm
\end{example}

\subsection{Maximally Vicious Computations}

We now prove that the devil's rules will cause infinite computations or failed states when these devil's rules and their recursive rule are applied alternatingly.
For the proof we need the following lemmata.
\begin{lemma}\label{disjointimplied}
{\rm
Given goal $C$ consisting of built-in constraints only and a goal $H$ consisting of user-defined constraints only.
Let the pairs $(H,C)$ with variables $x$ and $(H',C')$ with variables $y$ be disjoint variants.
Then the applicability condition
\[CT \models \forall \bar x (C \to \exists \bar y (H'{=}H \land C'))\]
trivially holds.

\medskip
\noindent {\bf Proof.}
Since $H$ and $H'$ are disjoint variants, the syntactic equality $H'{=}H$ is satisfiable. It implies a variable renaming between the variables in $x$ and $y$ that occur in $H$ and $H'$, respectively. We apply this variable renaming, replacing variables in $y$ in the applicability condition by the corresponding variables in $x$. 
This can only affect the consequent $(H'{=}H \land C')$ of the condition, where the constraints with the variables $y$ occur. 

In particular, the variable renaming will turn the equality $H'{=}H$ into $H{=}H$. Since this trivially holds, we can remove the equality. Moreover, the variable renaming will replace variables in $C'$ by their corresponding variables in $C$. Since $C$ and $C'$ are disjoint variants, the variables from $H$ will occur in the same positions in both expressions. Thus the two expressions will still be variants after the variable replacement.

This means that the premise and conclusion of the resulting implication can be written as
$\forall \bar w, \bar x' ( C[\bar x, \bar x'] \to \exists \bar y' C'[\bar x, \bar y'])$,
where $\bar x = \bar w, \bar x'$ and $\bar x'$ are the variables from $C$ and $\bar y'$ are the variables from $C'$ (and thus $y$) that have not been replaced.
This implication is logically equivalent to
$\forall \bar x (\exists \bar x' C[\bar x, \bar x'] \to \exists \bar y' C'[\bar x, \bar y'])$,
which is a tautology in first order predicate logic.
\qed
} %rm
\end{lemma}

\begin{lemma}\label{scheme}
{\rm
Given a self-recursive rule $r$ of the form
\[H_* \simp C_* \; \com \; B_{*bi} \land B_{*ud},\]   
and its devil's rule $d$ of the form
\[O_{B_{ud}} \setminus R_{B_{ud}} \simp C \land B_{bi} \; \com \; C' \land O_{B_{ud}}{=}O_{H'} \land R_{H'}.\]
Note that $B_{ud}{=}(O_{B_{ud}}\land R_{B_{ud}})$ and $H'{=}(O_{B_{ud}}\land R_{H'})$ since $O_{B_{ud}}{=}O_{H'}$.

Then according to the abstract semantics of CHR, any transition with $r$ and then $d$ has the form
   \[(H \land G) \mapsto_r \]% (B_{*bi} \land B_{*ud} \land  G \land H_*{=}H \land C_*)\]
\[(B_{*bi} \land B_{*ud} \land G \land H_*{=}H \land C_*) \mapsto_d\]

$\begin{array}{l}
((C' \land O_{B_{ud}}{=}O_{H'} \land H') \ \land \\
	\indent			B_{*bi} \land  G \land H_*{=}H \land C_* 
				\land (B_{ud}{=}B_{*ud} \land C \land B_{bi}))
\end{array}$
\medskip

\noindent with  $CT \models \forall (G_{bi} \to \exists \bar x (H_*{=}H \land C_*))$,
where $\bar x$ are the variables of a disjoint variant of the rule $r$,
and
with  $CT \models \forall ((B_{*bi} \land G_{bi} \land H_*{=}H \land C_*) \to \exists \bar y (B_{ud}{=}B_{*ud} \land C \land B_{bi}))$,
where $\bar y$ are the variables of a disjoint variant of the devil's rule $d$.

We will refer to the above transitions as {\em transition scheme}.
}
\end{lemma}

\begin{definition}\label{defmaxvicious}
{\rm
A {\em maximally vicious computation} of the self-recursive simplification rule $r$  and its devil's rules $D$ 
is of the form
\[(S'_{0}{=}(H_r \land C_r)) \ \ S'_{0} \mapsto_{r} T'_0 \ldots\ S'_i \mapsto_{r} T'_i \mapsto_{D} S'_{i+1} \ldots  (i \geq 0),\]
where the pair $(H_r, C_r)$ is a disjoint variant of the head and guard of the rule $r$.
}
\end{definition}

In a maximally vicious computation, the only user-defined constraints contained in the states 
come from the recursive rule.
\begin{lemma}\label{onlyHB}
{\rm
Given a maximally vicious computation of a recursive rule $r$ 
with head $H$ and body $B_{bi} \land B_{ud}$
and one of its devil's rules $d$.

Then the states $S_i$ contain as the only user-defined constraints a disjoint variant of the head $H$,
and the states $T_i$ contain as the only user-defined constraints a disjoint variant of the body $B_{ud}$.

\medskip
\noindent {\bf Proof.} 
By Definition \ref{defmaxvicious}, the first state $S_0$ contains the user-defined constraints $H_r$, which are a disjoint variant of the head of the rule $r$.
Consider the transition scheme of Lemma \ref{scheme}. For our inductive proof assume that $H$ is a disjoint variant of the head of the rule $r$ and that $G$ only contains built-in constraints, i.e. $G = G_{bi}$. 
An application of rule $r$ replaces user-defined constraints $H$ by $B_{*ud}$.
An application of a devil's rule $d$ of $r$ replaces user-defined constraints $B_{*ud}$ by $H'$, which is a disjoint variant of the head of rule $r$ by Definition \ref{devilsrule}.
\qed}
\end{lemma}

\begin{theorem}\label{maxvicious}
All maximally vicious computations of a self-recur\-sive simplification rule $r$ and its devil's rules $D$ 
are either non-termin\-at\-ing or end in a failed state.

{\rm
\medskip
\noindent {\bf Proof.} 
We prove the claim by induction over the computation steps.
The base case consists of showing that for any devil's rule $d$ in $D$ there exists a computation
\[(H_r \land C_r) = S'_{0} \mapsto_{r} T'_0 \mapsto_{d} S'_{1},\]
or a prefix of this computation ending in a failed state.
The induction step consists of two cases that together prove the claim:
\begin{enumerate}
\item If the self-recursive rule $r$ has been applied in a computation
and the resulting state is not failed, 
a devil's rule $d$ from $D$ associated with $r$ is applicable in the next computation step, i.e.
\[(i \geq 0) \land (S'_i \mapsto_{r} T'_i) \land \exists (T'_{ibi}) \to (T'_i \mapsto_{D} S'_{i+1}).\]
\item If a devil's rule $d$ from $D$ has been applied in a computation
and the resulting state is not failed, 
the self-recursive rule $r$ associated with it is applicable in the next computation step, i.e.
\[(i \geq 0) \land (T'_i \mapsto_{D} S'_{i+1}) \land \exists (S'_{i+1bi}) \to (S'_{i+1} \mapsto_{r} T_{i+1}).\]
\end{enumerate}

\medskip
\noindent {\bf Base Case.} 
According to the abstract semantics of CHR and Lemma \ref{scheme}, 
when we apply the direct recursive rule $r$ of the form 
\[r: H_* \simp C_* \; \com \; B_{*bi} \land B_{*ud},\]   
to the state $S'_{0} = (H_r \land C_r)$,
the transition is
    \[(H_r \land C_r) \mapsto_r (B_{*bi} \land B_{*ud} \land  C_r \land H_*{=}H_r \land C_*)\]
with  $CT \models \forall (C_r \to \exists \bar x (H_*{=}H_r \land C_*))$,
where $\bar x$ are the variables of the rule $r$.
By construction according to Definition \ref{devilsrule},
the pairs $(H_r, C_r)$ and $(H_*, C_*)$ are disjoint variants. 
Thus we can apply Lemma \ref{disjointimplied} to show that this applicability condition trivially holds.

If the target state is failed, then we are done with the proof of this case. 
Otherwise we apply to $B_{*ud}$ in the target state a devil's rule $d$ of the rule $r$ of the form
\[d: O_{B_{ud}} \setminus R_{B_{ud}} \simp C \land B_{bi} \; \com \; C' \land O_{B_{ud}}{=}O_{H'} \land R_{H'}.\]

This yields the transition
 \[(B_{*bi} \land B_{*ud} \land C_r \land H_*{=}H_r \land C_*) \mapsto_d\]

$\begin{array}{l}
(O_{B_{ud}} \land (C' \land O_{B_{ud}}{=}O_{H'} \land R_{H'}) \ \land \\
	\indent				B_{*bi} \land  C_r \land H_*{=}H_r \land C_* 
				\land (B_{ud}{=}B_{*ud} \land C \land B_{bi}))
\end{array}$
\medskip

\noindent with  $CT \models \forall (G'_{bi} \to \exists \bar y (B_{ud}{=}B_{*ud} \land C \land B_{bi}))$,
where $\bar y$ are the variables of the devil's rule $d$.
The built-in constraints $G'_{bi}$ of the source state are
$(B_{*bi} \land C_r \land H_*{=}H_r \land C_*)$,
and thus the following applicability condition must hold
\[CT \models \forall ((B_{*bi} \land C_r \land H_*{=}H_r \land C_*) \to \exists \bar y (B_{ud}{=}B_{*ud} \land C \land B_{bi})).\]
To show that this condition holds, it suffices to show that 
\[CT \models \forall ((C_* \land B_{*bi}) \to \exists \bar y (B_{ud}{=}B_{*ud} \land C \land B_{bi})).\]
By Definition \ref{devilsrule}, 
the tuples
$(B_{ud}, B_{bi}, C)$ and $(B_{*ud}, B_{*bi}, C_*)$ are disjoint variants. 
Thus we can apply Lemma \ref{disjointimplied} to show that this applicability condition trivially holds.

\medskip
\noindent {\bf Induction Step Case 1.} 
According to the abstract semantics of CHR and Lemma \ref{scheme}, 
when we apply the direct recursive rule $r$ of the form 
\[r: H_* \simp C_* \; \com \; B_{*bi} \land B_{*ud},\]   
the transition is
    \[({H} \land G) \mapsto_r (B_{*bi} \land B_{*ud} \land  G \land H_*{=}H \land C_*)\]
with  $CT \models \forall (G_{bi} \to \exists \bar x (H_*{=}H \land C_*))$,
where $\bar x$ are the variables of the rule $r$ and $G_{bi}$ are the built-in constraints in $G$.

If the target state is failed, then we are done with the proof of this case. 
Otherwise we apply to $B_{*ud}$ in the target state a devil's rule $d$ of the rule $r$ of the form
\[d: O_{B_{ud}} \setminus R_{B_{ud}} \simp C \land B_{bi} \; \com \; C' \land O_{B_{ud}}{=}O_{H'} \land R_{H'}.\]

This yields the transition
 \[(B_{*bi} \land B_{*ud} \land G \land H_*{=}H \land C_*) \mapsto_d\]

$\begin{array}{l}
 (O_{B_{ud}} \land (C' \land O_{B_{ud}}{=}O_{H'} \land R_{H'}) \ \land \\
	\indent			B_{*bi} \land  G \land H_*{=}H \land C_* 
				\land (B_{ud}{=}B_{*ud} \land C \land B_{bi}))
\end{array}$
\medskip

\noindent with  $CT \models \forall (G'_{bi} \to \exists \bar y (B_{ud}{=}B_{*ud} \land C \land B_{bi}))$,
where $\bar y$ are the variables of the devil's rule $d$.
The built-in constraints $G'_{bi}$ of the source state are
$(B_{*bi} \land G_{bi} \land H_*{=}H \land C_*)$,
and thus the following applicability condition must hold
\[CT \models \forall ((B_{*bi} \land G_{bi} \land H_*{=}H \land C_*) \to \exists \bar y (B_{ud}{=}B_{*ud} \land C \land B_{bi})).\]
To show that this condition holds, it suffices to show that 
\[CT \models \forall ((C_* \land B_{*bi}) \to \exists \bar y (B_{ud}{=}B_{*ud} \land C \land B_{bi})).\]
By Definition \ref{devilsrule}, %! Definition \ref{generalnonterm}, 
the tuples
$(B_{ud}, B_{bi}, C)$ and $(B_{*ud}, B_{*bi}, C_*)$ are disjoint variants. 
Thus we can apply Lemma \ref{disjointimplied} to show that this applicability condition trivially holds.

\medskip
\noindent {\bf Induction Step Case 2.} 
In a similar way we now prove the second claim.

Any devil's rule $d$ for the rule $r$ is of the form
\[d: O_{B_{ud}} \setminus R_{B_{ud}} \simp C \land B_{bi} \; \com \; C' \land O_{B_{ud}}{=}O_{H'} \land R_{H'}\]
It yields the transition 
\[(H \land G) \mapsto_{d}\] 
\[(O_{B_{ud}} \land (C' \land O_{B_{ud}}{=}O_{H'} \land R_{H'}) \land G
				\land (B_{ud}{=}H \land C \land B_{bi}))\]
with  $CT \models \forall (G_{bi} \to \exists \bar y (B_{ud}{=}H \land C \land B_{bi}))$,
where $\bar y$ are the variables of the devil's rule $d$,
and $G_{bi}$ are the built-in constraints in $G$.

If the target state is failed, then we are done with the proof of this case. 
Otherwise we apply to the target state the direct recursive rule $r$ of the form 
\[r: H_* \simp C_* \; \com \; B_{*bi} \land B_{*ud}.\]   
Since $O_{B_{ud}}{=}O_{H'}$, we can replace $O_{B_{ud}}$ by $O_{H'}$ in the target state.
But then we can replace $O_{H'} \land R_{H'}$ by $H'$ and apply rule $r$ to it.
The resulting transition is:
\[((C' \land O_{B_{ud}}{=}O_{H'} \land H') \land G \land (B_{ud}{=}H \land C \land B_{bi})) \mapsto_r\]

$\begin{array}{l}
((B_{*bi} \land B_{*ud}) \land  (C' \land O_{B_{ud}}{=}O_{H'}) \ \land \\
	\indent G \land (B_{ud}{=}H \land C \land B_{bi}) \land (H_*{=}H' \land C_*))
\end{array}$
\medskip

\noindent provided the following applicability condition holds

\medskip
$\begin{array}{l}
CT \models \forall (((C' \land O_{B_{ud}}{=}O_{H'}) \ \land \\
	\indent G_{bi} \land (B_{ud}{=}H \land C \land B_{bi})) \to \exists \bar x (H_*{=}H' \land C_*)),
\end{array}$
\medskip

\noindent where $\bar x$ are the variables of the rule $r$.

It suffices show that
$\forall (C' \to \exists \bar x (H_*{=}H' \land C_*))$.
The tuples $(H_*, C_*)$ and $(H',C')$ are disjoint variants.
Thus by Lemma \ref{disjointimplied} it is a tautology.
\qed}\end{theorem}

\subsection{Characterizing Non-Terminating Computations}

The next theorem shows that any non-terminating computation through a recursive rule in any program contains a maximally vicious computation of that recursive rule and its devil's rules. So if there is no non-terminating maximally vicious computation, then the recursive rule must be always terminating, no matter in which program it occurs.

We need the following two lemmata
from \cite{AbdennadherFruehwirth99}.
\begin{lemma}\label{lredund}
{\rm
A computation can be repeated in a state where
implied (or: redundant) built-in constraints have been removed.
Let $CT \models \forall \; (D \to C)$.%
\[\mbox{If }
(H \land C \land D
    \land G)
    \mapsto^* S
    \mbox{ then }
(H \land D \land G) \mapsto^* S.\]
}\end{lemma}

The next lemma states an important monotonicity property of CHR.
\begin{lemma}\label{lmonoton}
{\rm
{\em (CHR monotonicity)}
A computation can be repeated in any larger
context, i.e.\ with states in which built-in and user-defined constraints have been
added.
\[\mbox{If } G%_\glob 
\mapsto^* G'%_\glob
  \mbox{ then }(G \land H)%_{\glob'} 
\mapsto^* (G' \land H)%_{\glob'}
.\]
}
\end{lemma}

The following definition gives a necessary, sufficient, and decidable criterion for
equivalence of states~\cite{Raiser2009a}. %! Am I co-author?
\begin{definition}
{\rm
Given two states $S_1 = (S_{1bi} \land S_{1ud})$ and $S_2 = (S_{2bi} \land S_{2ud})$.
Then the two states are {\em equivalent}, written $S_1 \equiv S_2$, if and only if

\medskip
$\begin{array}{l}
	CT \models 
        \forall (S_{1bi} \rightarrow \exists \bar y ((S_{1ud} = S_{2ud}) \land S_{2bi}))
	\ \land \\
	\indent \forall (S_{2bi} \rightarrow \exists \bar x ((S_{1ud} = S_{2ud}) \land S_{1bi}))
\end{array}$
\medskip

\noindent with $\bar x$ those variables that only occur in $S_1$ and $\bar y$ those variables that only occur in $S_2$.
}
\end{definition}
Note that this notion (operational) equivalence is stricter than logical equivalence since it rules out idempotence of conjunction, i.e. it considers multiple occurrences of user-defined constraints to be different.

The overlap makes sure that the recursive rule is applied in a directly recursive way, 
common constraints of the overlap are denoted by $O_i$.
\begin{definition}\label{generalnonterm}
{\rm
A {\em non-terminating computation through a direct recursive rule $r$} 
\[H_* \simp C_* \; \com \; B_{*bi} \land B_{*ud},\]
in a program $P$ is of the form
\[S \mapsto_{P}^{n} S_{0} \ldots\ S_i \mapsto_{r} T_i \mapsto_{P}^{n_i} S_{i+1} \ldots \ \ (i \geq 0,\ n,n_i\geq 0),\]
where there is an overlap at common recursive constraints $O_i$ of $r$
with $T_i = (T_{Ri} \land O_i)$ and $S_{i+1} = (S_{Ri+1} \land O_i)$ %!!! \equiv?
where 
$O_i$ had been added by $r$ in the transition $S_i \mapsto_{r} T_i$
and $O_i$ is to be removed by $r$ in the transition $S_{i+1} \mapsto_{r} T_{i+1}$.

}
\end{definition}

\begin{theorem}\label{devil}
Given a self-recursive rule $r$ of the form
\[H_* \simp C_* \; \com \; B_{*bi} \land B_{*ud},\]   
and its devil's rules $d$ in $D$ of the form
\[O_{B_{ud}} \setminus R_{B_{ud}} \simp C \land B_{bi} \; \com \; C' \land O_{B_{ud}}{=}O_{H'} \land R_{H'}.\]
Given a non-terminating computation through $r$ in $P$, 
\[S \mapsto_{P}^{n} S_{0} \ldots\ S_i \mapsto_{r} T_i \mapsto_{P}^{n_i} S_{i+1} \ldots \ \ (i \geq 0,\ n,n_i\geq 0).\]
Then there is a corresponding maximally vicious % definition maxvicious exists
computation with $r$ and its devil's rules $D$
\[(S'_{0} = (H_r \land C_r)) \ \ S'_{0} \mapsto_{r} T'_0 \ldots\ S'_i \mapsto_{r} T'_i \mapsto_{D} S'_{i+1} \ldots \ \ (i \geq 0),\]
where there exist constraints $Gi$ such that 
\[S'_i \land G'_i \equiv S_i \ {\rm and} \ T'_i \land G'_i \equiv T_i,\]
and there exist overlaps at the common constraints $O_i$ that occur in the states
\[T'_i \equiv (T_{Ri} \land O_i) \ {\rm and} \ S'_{i+1} \equiv (S_{Ri+1} \land O_i).\]

{\rm
\medskip
\noindent {\bf Proof.}
We prove the claim by induction over the transitions in the computation.
The base case is to prove that $S'_0 \land G'_0 \equiv S_0$ and $T'_0 \land G'_0 \equiv T_0$
and the induction step means to prove
$S'_i \land G'_i \equiv S_i$ and $T'_i \land G'_i \equiv T_i$.

\medskip
\noindent {\bf Base Case 1.}  There exists a goal $G'_0$ such that $(S'_0 \land G'_0) = S_0$.

Let $S_0 = (H \land G)$.
According to the transition scheme in Lemma \ref{scheme}, we know that
$CT \models \forall (G_{bi} \to \exists \bar x (H_*{=}H \land C_*))$,
where $\bar x$ are the variables of the rule $r$.
As defined by the CHR semantics, $(H \land G)$ does not contain variables from $\bar x$.
Let $S'_{0} = (H'_* \land C'_*)$, such that  % used H_r and C_r in stating Theorem above
$(H'_* \land C'_*)$ does not contain variables from $(H \land G)$ and $(H_* \land C_*)$.
Let $G'_0$ be $(H'_*{=}H \land G)$, then 
$(S'_0 \land G'_0) = ((H'_* \land C'_*) \land (H'_*{=}H \land G))$.

Since $H'_*{=}H$, we can replace the first conjunct $H'_*$ by $H$.
Since
$CT \models \forall (G_{bi} \to \exists \bar x (H_*{=}H \land C_*))$,
and since
$(H'_* \land C'_*)$ and $(H_* \land C_*)$ are disjoint variants,
by Lemma \ref{disjointimplied} it also holds that 
$CT \models \forall (G_{bi} \to \exists \bar z (H'_*{=}H \land C'_*))$,
where $z$ are the variables of $(H'_* \land C'_*)$.
Therefore, $(H'_*{=}H \land C'_*)$ is redundant and can be removed according to Lemma \ref{lredund}.
Hence $(S'_0 \land G'_0) = (H \land G) = S_0$.

\medskip
\noindent {\bf Base Case 2.}  Given $G'_0 = (H'_*{=}H \land G)$ from Base Case 1, 
we proceed to prove $(T'_0 \land G'_0) = T_0$.

W.l.o.g. we apply to $S'_0$ and $S_0$ the same disjoint variant of rule $r$, namely 
$H_* \simp C_* \; \com \; B_{*bi} \land B_{*ud}$, to reach $T'_0$ and $T_0$, respectively.

Then we have that
$T_0 = (B_{*bi} \land B_{*ud} \land G \land H_*{=}H \land C_*)$
and 
$(T'_0 \land G'_0) =
((B_{*bi} \land B_{*ud} \land C'_*  \land H_*{=}H'_*  \land C_*) \land (H'_*{=}H \land G))$.
In the state $(T'_0 \land G'_0)$ we can replace $H_*{=}H'_*$ by $H_*{=}H$, since the state also contains $H'_*{=}H$.
Now the states differ only in that $(H'_*{=}H \land C'_*)$ additionally occurs in $(T'_0 \land G'_0)$.
Analogously to the reasoning for the Base Case 1, we can show that this conjunction is redundant. 
So these states are indeed equivalent.

\medskip
\noindent {\bf Induction Step Case 1.}  
We prove that there exists a $G'_i$ such that
$S'_i \land G'_i \equiv S_i$.

According to the transition scheme in Lemma \ref{scheme} 
we know that any transition with $r$ and then some $d$ in any computation has the form
   \[(H \land G) \mapsto_r \] % (B_{*bi} \land B_{*ud} \land  G \land H_*{=}H \land C_*)\]
 \[(B_{*bi} \land B_{*ud} \land G \land H_*{=}H \land C_*) \mapsto_d\]

 $\begin{array}{l}
((C' \land O_{B_{ud}}{=}O_{H'} \land H') \ \land \\
	\indent           B_{*bi} \land  G \land H_*{=}H \land C_* 
				\land (B_{ud}{=}B_{*ud} \land C \land B_{bi}))
\end{array}$
\medskip

\noindent with  $CT \models \forall (G_{bi} \to \exists \bar x (H_*{=}H \land C_*))$,
where $\bar x$ are the variables of a disjoint variant of the rule $r$,
and
with  $CT \models \forall ((B_{*bi} \land G_{bi} \land H_*{=}H \land C_*) \to \exists \bar y (B_{ud}{=}B_{*ud} \land C \land B_{bi}))$,
where $\bar y$ are the variables of a disjoint variant of the devil's rule $d$.

In a maximally vicious computation, by Lemma \ref{onlyHB} we know that $G$ above does not contain user-defined constraints, i.e. $G=G_{bi}$.

W.l.o.g. let $S'_i$ be the last state of the above transitions. 

Furthermore, any transition with $r$ and then $P$ in any computation has the form
   \[(H \land G'_{i-1}) \mapsto_r \] % (B_{*bi} \land B_{*ud} \land  G_{bi} \land H_*{=}H \land C_*)\]
 \[(B_{*bi} \land B_{*ud} \land G'_{i-1} \land H_*{=}H \land C_*) \mapsto^{n_i}_P\]
 \[(B_{*bi}               \land G''_i \land H_*{=}H \land C_*).\]

W.l.o.g. let $S_i$ be $(B_{*bi} \land G''_i \land H_*{=}H \land C_*)$.
Since $r$ is applicable to the target state $S_i$, 
$G''_i$ must contain user-defined constraints $H'_i$ such that 
\[CT \models \forall ((B_{*bi} \land G''_{ibi} \land H_*{=}H \land C_*) \to \exists \bar x (H_{i*}{=}H'_i \land C_{i*})),\]
where $\bar x_i$ are the variables of a disjoint variant of rule $r$ with head $H_{i*}$ and guard $C_{i*}$.

By Definition \ref{devilsrule} and Definition \ref{generalnonterm},
$H'_i$ must overlap with $B_{*ud}$.
This overlap at the common constraints $O_{H'_i}$ in $H'_i$ of state $S_i$ has its 
correspondence in the overlap at the common constraints $O_{B_{*ud}}$ in $B_{*ud}$ of state $T_i$. 
This means it must hold that $CT \models \forall (G''_{ibi} \to \exists (O_{B_{*ud}}{=}O_{H'_i}))$.

The previous state $T'_{i-1}$ contains $G_{bi}$.
Since $T'_{i-1}$ is contained in $T_{i-1}$, it must be (implied) there as well.
According to the CHR semantics, built-in constraints are accumulated during a computation.
Thus it must hold that $CT \models \forall (G''_{ibi} \to \exists G_{bi})$.

In the previous state $T'_{i-1}$ it holds that
\[CT \models \forall ((B_{*bi} \land G_{bi} \land H_*{=}H \land C_*) \to \exists \bar z (B_{ud}{=}B_{*ud} \land C \land B_{bi})),\] 
where $\bar z$ are the variables of
$(B_{ud} \land C \land B_{bi})$. Since $T'_{i-1}$ is contained in $T_{i-1}$, the implication must also hold there.
Since built-in constraints are accumulated, the implication must also hold in the next state $S_i$.

Putting these observation all together, we now know the state $S_i$ is of the form

\medskip
$\begin{array}{l}
((H'_i \land (H_{i*}{=}H'_i \land C_{i*}) \land O_{B_{*ud}}{=}O_{H'_i}) \land G_{bi} \land B_{*bi}\ \land \\
 \indent G'''_i \land H_*{=}H \land C_* 
				\land (B_{ud}{=}B_{*ud} \land C \land B_{bi})),
\end{array}$
\medskip

\noindent where $G'''_i$ are the remaining constraints from $G''_i$.
Since $B_{ud}{=}B_{*ud}$, we can write $O_{B_{*ud}}{=}O_{H'_i}$ as $O_{B_{ud}}{=}O_{H'_i}$.
Since $(H_{i*},C_{i*})$ and $(H',C')$ are disjoint variants, we can apply Lemma \ref{disjointimplied} to show that
$CT \models \forall (C_{i*} \to \exists (H'{=}H_{i*} \land C'))$.
Finally, let $G'_i = G'''_i$, then $S'_i \land G'_i \equiv S_i$.

\medskip
\noindent {\bf Induction Step Case 2.}  
Finally, we prove that $(T'_i \land G'_i) = T_i$.

We know that $(S'_i \land G'_i) = S_i$.
Since $S'_i \mapsto_r T'_i$, by CHR monotonicity (Lemma \ref{lmonoton}) we have that 
$S'_i \land G'_i \mapsto_r T'_i \land G'_i$.
Thus $T_i = (T'_i \land G'_i)$.
\qed}\end{theorem}

\begin{example}\label{even23} %! user tabular or array format layout
{\rm 
We continue with Example \ref{even2}, its recursive rule and its simplified devil's rule
\[r: \mathit{even(X)} \simp X{=}s(Y) \; \com \; Y{=}s(Z) \land \mathit{even(Z)}\]
\[d: \mathit{even(X)} \prop X{=}s(Y).\]

For readability, we will simplify states w.r.t. equivalence $\equiv$
and underline the goals to which a rule is applied.

The maximally vicious computation is
\[\underline{\mathit{even(U)} \land U{=}s(V)} \mapsto_{r}\]
\[
U{=}s(V) \land V{=}s(Z') \land \underline{\mathit{even(Z')}} \mapsto_{d}\]
\[U{=}s(V) \land V{=}s(Z') \land \underline{\mathit{even(Z')} \land Z'{=}s(Y'')} \mapsto_{r} \ldots\]

The non-terminating computation for the goal $\mathit{even(N)} \land \mathit{even(M)} \land M{=}s(N)$ is
\[{\rm{even(N)}} \land \underline{\mathit{even(M)} \land M{=}s(N)} \mapsto_{r}\]
\[\underline{{\rm{even(N)}}} \land M{=}s(N) \land \underline{N{=}s(N')} \land \mathit{even(N')} \mapsto_{r}\]
\[N{=}s(N') \land \underline{N'{=}s(N'')} \land {\rm{even(N'')}}   \land M{=}s(N) \land \underline{\mathit{even(N')}} \mapsto_{r}{..}\]
This computation contains the maximally vicious computation when we rename the variables appropriately. 
Note that the second application of the rule $r$ to the other, {\em even(N)} constraint
is considered as the arbitrary transition sequence between recursive steps using rules from a given program $P$.
The actual infinite computation differs from the maximally vicious computation only in an additional {\em even} constraint 
(that we have set in standard font style).
} %rm
\end{example}

\section{Static Termination and Non-Termination Analysis with Devil's Rules}

We identify cases where the static analysis of the devil's rules decides termination or non-termination of the recursive rule in {\em any} program as exhibited by a maximally vicious computation.
These conditions are just meant to be indicative of the potential of our devil's advocate approach, they are a starting point in the search for interesting conditions.
The first lemma gives a necessary condition for non-termination. 
The negation of the condition thus gives a sufficient condition for termination.
The second lemma gives a sufficient condition for non-termination. It implies the first condition, but it is not a sufficient and necessary condition.

\subsection{A Termination Condition}

\begin{lemma}\label{nonfailed}
{\rm
Given a recursive rule $r$ and a devil's rule $d$ for $r$ of the form
\[d: O_{B_{ud}} \setminus R_{B_{ud}} \simp C \land B_{bi} \; \com \; C' \land O_{B_{ud}}{=}O_{H'} \land R_{H'}\]
Then the computation 
\[(H_r \land C_r) = S'_{0} \mapsto_{r} T'_0 \mapsto_{d} S'_{1}\]
or any of its prefixes does not end in a failed state,
if and only if
\[C \land B_{bi} \land C' \land O_{B_{ud}}{=}O_{H'}\]
is consistent.

\medskip
\noindent {\bf Proof.} 
From Lemma \ref{scheme} and the proof of Theorem \ref{maxvicious}
we can see that the built-in constraints of the three states in the computation are
\[S'_{0bi} = (C_r)\]
\[T'_{0bi} = ((B_{*bi}) \land C_r \land (H_*{=}H_r \land C_*))\]

$\begin{array}{l}
S'_{1bi} = ((C' \land O_{B_{ud}}{=}O_{H'}) \ \land \\
	\indent		B_{*bi} \land  C_r \land H_*{=}H_r \land C_* 
				\land (B_{ud}{=}B_{*ud} \land C \land B_{bi})),
\end{array}$
\medskip

\noindent where constraints in brackets are newly added in a state.
We also know that the tuples
$(H_r, C_r)$ and $(H_*, C_*)$ and $(H',C')$ as well as
$(B_{ud}, B_{bi}, C)$ and $(B_{*ud}, B_{*bi}, C_*)$  
are disjoint variants and that
the constraints 
$(H_*{=}H_r \land C_*)$ and $(B_{ud}{=}B_{*ud} \land C \land B_{bi})$
are implied by the states $S'_{0bi}$ and $T'_{0bi}$, respectively.

According to the semantics of CHR, built-in constraints in a computation are accumulated, i.e. added but never removed. 
To prove the claim, it therefore suffices to consider the built-in constraints of the last state $S'_{1bi}$.
We have to show that

\medskip
$\begin{array}{l}
C \land B_{bi} \land C' \land O_{B_{ud}}{=}O_{H'} \ \leftrightarrow \\
	\indent	\exists \bar x (C' \land O_{B_{ud}}{=}O_{H'}   \land
				B_{*bi} \land  C_r \land H_*{=}H_r \land C_* \ \land \\
	\indent \indent 	B_{ud}{=}B_{*ud} \land C \land B_{bi}),
\end{array}$
\medskip

\noindent where $\bar x$ consist of all variables that do not occur in the left-hand side of the logical equivalence.
Since the left-hand side constraints occur in the right hand side, we just have to show that the left-hand side implies the right-hand side. 
As in the the proof of Theorem \ref{maxvicious}, we can apply Lemma \ref{disjointimplied} to show that
remaining constraints on the right-hand side 
$B_{*bi} \land  C_r \land H_*{=}H_r \land C_* \land B_{ud}{=}B_{*ud}$ 
are implied by the left-hand side, since they are disjoint variants as given above.
\qed}
\end{lemma}
Note that the lemma does not say that any computation with the recursive rule alone will fail. 
But it will fail with a transition of the devil's rule.

\begin{example}\label{even3} %! user tabular or array format layout
{\rm
We continue with Example \ref{even23} and its %rule
correct devil's rule
\[\mathit{even(Z)}{\Rightarrow}X{=}s(Y) \land Y{=}s(Z) \com  X'{=}s(Y') \land \mathit{even(Z){=}even(X')}.\]
The conjunction to check can be simplified into
\[X{=}s(Y) \land Y{=}s(Z) \land Z{=}s(Y') \land X'{=}Z\]
and is clearly satisfiable.
Thus non-termination is not ruled out.
} %rm
\end{example}

\begin{example}\label{c01} %! user tabular or array format layout
{\rm 
Consider the direct recursive rule and its devil's rule
\[c(0) \simp c(s(X)).\]
\[c(s(X)) \prop c(s(X)){=}c(0).\]
Since $s(X){=}0$ is unsatisfiable, the recursive rule must always terminate.
Actually, the recursive goal $c(s(X))$ will delay.
} %rm
\end{example}

\subsection{A Non-Termination Condition}

\begin{lemma}\label{non-terminating}
{\rm
Given a recursive rule $r$ and a devil's rule $d$ for $r$ of the form
\[d: O_{B_{ud}} \setminus R_{B_{ud}} \simp C \land B_{bi} \; \com \; C' \land O_{B_{ud}}{=}O_{H'} \land R_{H'}\]
and let $(H, C, B_{bi})$ and $(H', C', B'_{bi})$ be disjoint variants derived from the recursive rule $r$.
Then the maximally vicious computation 
\[(S'_{0} = (H_r \land C_r)) \ \ S'_{0} \mapsto_{r} T'_0 \ldots\ S'_i \mapsto_{r} T'_i \mapsto_{d} S'_{i+1} \ldots
 \ \ (i \geq 0)\]
is non-terminating,
if 
\[{\mathit (NT)} \ \ \exists (C \land B_{bi}) \land 
                    \forall ((C \land B_{bi}) \to \exists (C' \land O_{B_{ud}}{=}O_{H'} \land B'_{bi})).\]

\medskip
\noindent {\bf Proof.}
We prove by induction over the transitions analgous to Theorem \ref{maxvicious}. 
We show that if a state in the computation is not failed and condition {\em NT} holds, then the next state is not failed as well. 
In the induction step we distinguish between applications of rule $r$ and rule $d$.

\medskip
\noindent {\bf Base Case.}
For the base case, we consider the prefix of the computation
\[(H_r \land C_r) = S'_{0} \mapsto_{r} T'_0 \mapsto_{d} S'_{1}.\]
Condition {\em NT} implies by the laws of first-order predicate logic
\[\exists (C \land B_{bi} \land C' \land O_{B_{ud}}{=}O_{H'}).\]
From Lemma \ref{nonfailed} we know that this conjunction implies that the prefix of the computation has no failed states.

\medskip
\noindent {\bf Induction Step Case 1.} The transition for rule $r$ yields the condition
\[(i \geq 0) \land {\mathit NT} \land \exists (S'_i) \land (S'_i \mapsto_{r} T'_i) \to \exists (T'_{ibi}).\]
According to Lemma \ref{scheme} and Theorem \ref{maxvicious}, this transition is of the form
\[((C' \land O_{B_{ud}}{=}O_{H'} \land H') \land G \land (B_{ud}{=}H \land C \land B_{bi})) \mapsto_r\]

$\begin{array}{l}
((B_{*bi} \land B_{*ud}) \land  (C' \land O_{B_{ud}}{=}O_{H'}) \ \land \\
	\indent	G \land (B_{ud}{=}H \land C \land B_{bi}) \land (H_*{=}H' \land C_*)).\\
\end{array}$
\medskip

In the target state, the built-in constraints $(C' \land O_{B_{ud}}{=}O_{H'}) \land G_{bi} \land (B_{ud}{=}H \land C \land B_{bi})$ are satisfiable, because they already occured in the source state.
The new constraints $(H_*{=}H' \land C_*)$ are satisfiable, 
because they are implied by $(C')$ in the source state (due to the applicability condition that must hold for the transition). 
They have been added by the rule application together with $B_{*bi}$.

The constraints $(C \land B_{bi})$ of the source state must be satisfiable and by condition {\em NT} we have that
\[\forall (C \land B_{bi} \to \exists (C' \land O_{B_{ud}}{=}O_{H'} \land B'_{bi})).\]
The tuples $(H', C', B'_{bi})$ and $(H_*, C_*, B_{*bi})$ are disjoint variants. 
Therefore $(C_* \land B_{*bi})$ must be consistent as well, 
and thus the built-in constraints of the target state are all satisfiable.

\medskip
\noindent {\bf Induction Step Case 2.} The transition for a devil's rule $d$ yields the condition
\[(i \geq 0) \land {\mathit NT} \land \exists (T'_i) \land (T'_i \mapsto_{d} S'_{i+1}) \to \exists (S'_{i+1bi}).\]
According to Lemma \ref{scheme} and Theorem \ref{maxvicious}, this transition is of the form
 \[(B_{*bi} \land B_{*ud} \land G \land H_*{=}H \land C_*) \mapsto_d\]

$\begin{array}{l}
(O_{B_{ud}} \land (C' \land O_{B_{ud}}{=}O_{H'} \land R_{H'}) \ \land \\
    \indent			B_{*bi} \land  G \land H_*{=}H \land C_* 
				\land (B_{ud}{=}B_{*ud} \land C \land B_{bi})).
\end{array}$
\medskip

In the target state, the constraints $(B_{*bi} \land B_{*ud} \land G_{bi} \land H_*{=}H \land C_*)$ are satisfiable, because they occur in the source state.
The new constraints $(B_{ud}{=}B_{*ud} \land C \land B_{bi})$ are satisfiable, 
because they are implied by $(C_* \land B_{*bi})$ in the source state (due to the applicability condition that must hold for the transition). 
They have been added by the rule application together with $(C' \land O_{B_{ud}}{=}O_{H'})$.
By condition {\em NT} we have that
\[\forall (C \land B_{bi} \to \exists (C' \land O_{B_{ud}}{=}O_{H'} \land B'_{bi})).\]
Therefore $(C' \land O_{B_{ud}}{=}O_{H'})$ must be consistent as well.
Thus the built-in constraints of the target state are all satisfiable.
\qed}
\end{lemma}
If condition {\em NT} holds for a devil's rule, its maximally vicious computation is non-terminating.
Thus there may be other non-terminating computations for the recursive rule. 
Conversely, if the maximally vicious computation terminates in a failed state, the condition cannot hold.
On the other hand, if the condition does not hold, we cannot draw any conclusion about non-termination from it.
We rather have to look at the maximally vicious computation for further insight about the termination behavior.

\begin{example}\label{prime} %! user tabular or array format layout
{\rm 
Let {\em odd} and {\em prime} be built-in constraints.   %! used only here
Consider the following recursive rule and its devil's rule
\[c(X) \simp \mathit{odd}(X)  \; \com \;  c(s(s(X))),\]
\[c(s(s(X))) \prop \mathit{odd}(X)  \; \com \;  (\mathit{odd}(X') \land c(s(s(X))){=}c(X')).\]
Condition {\it NT} amounts to
\[\exists \mathit{odd}(X)  \land \forall (\mathit{odd}(X) \to (\mathit{odd}(X') \land c(s(s(X))){=}c(X'))).\]
Since the successor of the successor of an odd number is always odd, the condition holds.
Actually, the recursive rule on its own is non-terminating for odd numbers. %!

Now consider a variation of the above rule
\[c(X) \simp \mathit{prime}(X)  \; \com \;  c(s(s(X))).\]
Condition {\it NT} amounts to
\[\exists \mathit{prime}(X)  \land \forall (\mathit{prime}(X){\to}(\mathit{prime}(X') \land c(s(s(X))){=}c(X'))).\]
Since the successor of the successor of a prime number may not be prime, the condition does not hold. 
Thus the status of non-termination is undecided. 
Actually, the recursive rule always terminates after at most two recursive steps: 
one of every three sequential odd numbers is a multiple of three, and hence not prime.
Hence the maximally vicious computation always ends in a failed state.
} %rm
\end{example}

\section{Examples - Putting It All Together}\label{examples}

\noindent {\bf Devil's Rule Simplification.}
In practice, we will simplify the built-in constraints in the devil's rules by replacing them with logically equivalent ones. 
We do so taking into account that variables not occurring in the head of a rule are implicitly existentially quantified according to the CHR semantics. 
Moreover, if built-in constraints of the body are implied by the guard, we can remove them if other constraints in the body are not affected.

We can then distinguish two extreme cases:
\begin{enumerate}
\item
If the built-in constraints in the devil's rule are inconsistent, 
we can replace the body of the rule by the built-in constraint $\mathit{false}$. We have a case for Lemma \ref{nonfailed}. 
So the recursive rule is unconditionally terminating if all its devil's rules simplify in this way. 

\item
If the simplification of a devil's rule yields a satisfiable guard and
a body built-in constraint equivalent to $\mathit{true}$, then Lemma \ref{non-terminating} may apply.
Part of the condition $\mathit{NT}$ of the lemma already holds in that case. 
It remains to check if the body built-in constraint of the recursive rule is implied in the context of the condition. 
This trivially holds if there are no such body constraints.
\end{enumerate}

\subsection{Even Numbers}

We continue with Example \ref{even3}. 
Recall the correct rule for {\em even} and its devil's rule
\[\mathit{even(X)} \simp X{=}s(Y) \; \com \; Y{=}s(Z) \land \mathit{even(Z)}.\]
\[\mathit{even(Z)}{\Rightarrow}X{=}s(Y) \land Y{=}s(Z)  \com  X'{=}s(Y') \land \mathit{even(Z){=}even(X')}.\]
The devil's rule can be simplified into $\mathit{even(Z)} \prop Z{=}s(Y')$.

Lemma \ref{nonfailed} does not apply.
Lemma \ref{non-terminating} yields the condition {\em NT} (with the equality $\mathit{even(Z){=}even(X')}$ simplified away for convenience)

\medskip
$\begin{array}{l}
\exists (X{=}s(Y) \land Y{=}s(Z)) \ \land \\
    \indent \forall ((X{=}s(Y) \land Y{=}s(Z)) \to \exists (Z{=}s(Y') \land Y'{=}s(Z'))).\end{array}$
\medskip

Since the predecessor of the predecessor of an even natural number does not exist for the number $0$, it is not always even. Thus the condition does not hold. 
But it easy to see from the recursive rule and its devil's rule that 
the maximally vicious computation does not terminate.
Actually, the recursive rule on its own may not terminate as we have shown in the introduction.

\medskip
\noindent {\bf Termination.}
Next we consider two examples for the application of Lemma \ref{nonfailed} for termination.
Consider an erroneous version of the rule for {\em even} with a typo (highlighted by bold type font)
\[\mathit{even(X)} \simp X{=}s(Y) \; \com \; Y{=}s(\bf{X}) \land \mathit{even(Z)}.\]
This leads to a devil's rule with the unsatisfiable guard $X{=}s(Y) \land Y{=}s(X)$. After rule simplification we arrive at
\[\mathit{even(Z)} \prop \mathit{false} \; \com \; \mathit{false}.\]
Since the body of the simplified devil's rule is $\mathit{false}$, the recursive rule will always terminate. It will actually lead to a failed state, when it is applicable, since its guard and body built-in constraints are in contradiction.

Now consider another typo and the resulting simplified devil's rule
\[\mathit{even(X)} \simp X{=}s(Y) \; \com \; Y{=}s(\bf{z}) \land \mathit{even(\bf{z})}.\]
\[\mathit{even(z)} \prop  \mathit{false}.\]
The recursive rule will thus always terminate, but not necessarily in a failed state. Once the recursive rule is applied, it cannot be applied again to the recursive goal $\mathit{even(z)}$, since the argument $z$ does not satisfy the guard that demands a term of the form $s(Y)$.

\medskip
\noindent {\bf Non-termination.}
Now we look at examples for application of Lemma \ref{non-terminating} for non-termination.
Consider the following rule with a typo and its devil's rule
\[\mathit{even(X)} \simp X{=}s(Y) \; \com \; Y{=}s(Z) \land \mathit{even(\bf{X})}.\]
\[\mathit{even(X)}{\Rightarrow}X{=}s(Y) \land Y{=}s(Z)  \com  X'{=}s(Y') \land \mathit{even(X){=}even(X')}.\]
After simplification we arrive at
\[\mathit{even(X)} \prop X{=}s(Y) \land Y{=}s(Z) \; \com \; \mathit{true}.\]
The implication in condition {\em NT} corresponds to 
\[\forall ((X{=}s(Y) \land Y{=}s(Z)) \to \exists (X{=}s(Y') \land Y'{=}s(Z'))).\]
Actually, the recursive rule on its own is already always non-terminating.

Now consider another erroneous rule
and the resulting simplified devil's rule
\[\mathit{even(X)} \simp X{=}s(Y) \; \com \; Y{=}s(\bf{z}) \land \mathit{even(\bf{Y})}.\]
\[\mathit{even(Y)} \prop  Y{=}s(z) \; \com \; \mathit{true}.\] 
The implication in condition {\em NT} corresponds to 
\[\forall ((X{=}s(Y) \land Y{=}s(z)) \to \exists (Y{=}s(Y') \land Y'{=}s(z))).\]
The condition does not hold, since $Y{=}s(z)$ and $Y{=}s(Y') \land Y'{=}s(z)$ are in contradiction.
So $\mathit{true}$ in the body of a simplified devil's rule does not necessarily mean non-termination, 
the maximally vicious computation may end in a failed state.
Actually, here any computation where the recursive rule is applied will lead to a failed state.

\subsection{Minimum}\label{gamma-min}

We compute the minimum of a multiset of numbers $n_i$, 
and $\leq$ a non-strict total order over an infinite domain of numbers, %!
given as a computation of the query
$\mathit{min}(n_1), \mathit{min}(n_2),..., \mathit{min}(n_k)$ with the recursive rule
\[\mathit{min}(N) \land \mathit{min}(M) \simp N \leq M \; \com \; \mathit{min}(N).\]
The rule takes two $\mathit{min}$ candidates and removes the one with the
larger value. It keeps going until only one, the smallest
value, remains as single $\mathit{min}$ constraint.

There are two overlaps at the recursive constraint {\em min} in the rule. 
The resulting devil's rules and then their simplified versions are

\medskip
$\begin{array}{l}
\mathit{min}(N) \prop N \leq M \; \com \; N' \leq M' \land \\
    \indent \mathit{min}(N){=}\mathit{min}(N') \land \mathit{min}(M')\\
\mathit{min}(N) \prop N \leq M \; \com \; N' \leq M' \land \\
    \indent \mathit{min}(N){=}\mathit{min}(M') \land \mathit{min}(N')
\end{array}$

\medskip
$\begin{array}{l}
\mathit{min}(N) \prop N \leq M' \land \mathit{min}(M')\\
\mathit{min}(N) \prop N' \leq N \land \mathit{min}(N')
\end{array}$
\medskip

These are propagation rules that either add a smaller or larger $min$ constraint.
According to Lemma \ref{non-terminating}, 
the condition ${\mathit NT}$ amounts to

\medskip
$\begin{array}{l}
\exists (N \leq M) \land \forall ((N \leq M) \to \exists (N \leq M'))\\
\exists (N \leq M) \land \forall ((N \leq M) \to \exists (N' \leq N))
\end{array}$
\medskip

Since both conditions hold, 
all maximally vicious computations are indeed non-terminating, 
no matter which of the two devil's rules are used.
Actually, the recursive rule on its own does not terminate if we keep adding $min$ constraints. 
Otherwise, it terminates, since every rule application removes one $min$ constraint.

\subsection{Exchange Sort}

We can \index{sort} sort an array by keeping exchanging values at positions that are in
the wrong order. 
Given an array as a conjunction of constraints representing array elements
$a(\mathit{Index,Value})$, i.e.  $a(1,A_1) \land \ldots \land a(n,A_n)$, 
and a strict total order $<$ over the integers, %! would work with any numbers
the following recursive rule sorts in this way
\[a(I,V) \land a(J,W) \simp I{>}J \land V{<}W \; \com \; a(I,W) \land a(J,V).\]
In a sorted array, it holds for each pair $a(I,V), a(J,W)$ where 
$I{>}J$ that $V \geq W$.  The rule ensures that this indeed will hold for
every such pair by exchanging the values if necessary.  

There are two full overlaps, the resulting devil's rules and their simplified versions are

\medskip
$\begin{array}{l}
a(I,W) \land a(J,V) \prop I{>}J \land V{<}W \; \com \; I'{>}J' \land V'{<}W' \land \\
    \indent (a(I,W) \land a(J,V)){=}(a(I',V') \land a(J',W'))\\
a(I,W) \land a(J,V) \prop I{>}J \land V{<}W \; \com \; I'{>}J' \land V'{<}W' \land \\
    \indent (a(I,W) \land a(J,V)){=}(a(J',W') \land a(I',V'))
\end{array}$

\medskip
$\begin{array}{l}
a(I,W) \land a(J,V) \prop I{>}J \land V{<}W \; \com \; \mathit{false}\\   % I=I' \land J=J' \land W=V' \land V=W'\\
a(I,W) \land a(J,V) \prop I{>}J \land V{<}W \; \com \; \mathit{false}\\   % I=J' \land J=I' \land V=V' \land W=W'
\end{array}$
\medskip

To the devil's rules of the full overlaps Lemma \ref{nonfailed} applies. 
Indeed, the recursive rule terminates for any two array constraints. 
It cannot be applied a second time to the same pair of constraints.

There are four more partial overlaps between one array constraint from the head and one from the body of the recursive rule, yielding four devil's rules and their simplifications

\medskip
$\begin{array}{l}
a(I,W) \setminus a(J,V) \simp I{>}J \land V{<}W \; \com \;\\
    \indent I'{>}J' \land V'{<}W' \land a(I,W){=}a(I',V') \land a(J',W')\\

a(I,W) \setminus a(J,V) \simp I{>}J \land V{<}W \; \com \;\\
    \indent  I'{>}J' \land V'{<}W' \land a(I,W){=}a(J',W') \land a(I',V')\\

a(J,V) \setminus a(I,W) \simp I{>}J \land V{<}W \; \com \;\\
    \indent  I'{>}J' \land V'{<}W' \land a(J,V){=}a(I',V') \land a(J',W')\\

a(J,V) \setminus a(I,W) \simp I{>}J \land V{<}W \; \com \;\\
    \indent  I'{>}J' \land V'{<}W' \land a(J,V){=}a(J',W') \land a(I',V')
\end{array}$

\medskip
$\begin{array}{l}

a(I,W) \setminus a(J,V) \simp I{>}J \land V{<}W \; \com \; \\
    \indent I{>}J' \land W{<}W' \land a(J',W')\\

a(I,W) \setminus a(J,V) \simp I{>}J \land V{<}W \; \com \; \\
    \indent I'{>}I \land V'{<}W \land a(I',V')\\

a(J,V) \setminus a(I,W) \simp I{>}J \land V{<}W \; \com \; \\
    \indent J{>}J' \land V{<}W' \land a(J',W')\\

a(J,V) \setminus a(I,W) \simp I{>}J \land V{<}W \; \com \; \\
    \indent I'{>}J \land V'{<}V \land a(I',V')
\end{array}$
\medskip

The condition ${\mathit NT}$ of Lemma \ref{non-terminating} is satisfied for these four devil's rules, for example consider

\medskip
$\begin{array}{l}
\exists (I{>}J \land V{<}W) \ \land \\
    \indent \forall ((I{>}J \land V{<}W) \to \exists (I{>}J' \land W{<}W')).
\end{array}$
\medskip

To cause non-termination, the devil's rules replace an array constraint by another one. For every pair of array elements that has been ordered, this replacement results in an unordered pair.
Therefore the array sort rule may not terminate if the array is updated during sorting.

We next consider an erroneous version of array sort, where the guard condition $I{>}J$ is missing, and its two full overlaps.

\medskip
$\begin{array}{l}
a(I,V) \land a(J,W) \simp V{<}W \; \com \; a(I,W) \land a(J,V)
\end{array}$

\medskip
$\begin{array}{l}
a(I,W) \land a(J,V) \prop V{<}W \; \com \; \mathit{false}\\   % I=I' \land J=J' \land W=V' \land V=W'
a(I,W) \land a(J,V) \prop V{<}W \; \com \; \mathit{true}\\   % I=J' \land J=I' \land V=V' \land W=W'
\end{array}$
\medskip

The second full overlap now produces a different simplified devil's rule. It has the body $\mathit{true}$.
Moreover, the recursive rule does not have any built-in constraints in its body. 
Therefore Lemma \ref{non-terminating} holds. 
Actually, the guard condition $V{<}W$ applies to any pair of array constraints with different values.
So the rule will keep exchanging values in such two array constraints forever.

\section{Conclusions}

In this paper we have introduced a novel approach to non-termination analysis, exemplified for the programming language Constraint Handling Rules (CHR).
It is based on the notion of a devil's advocate that produces a malicious program that causes non-termination of the given program, if it is possible at all.
We have introduced the devil's advocate method using direct recursive simplifications rules of CHR. From them, the devil's advocate constructs so-called devil's rules in a simple manner. If a recursive rule and its devil's rules are applied alternatingly, the result is a maximally vicious computation. It is either infinite or ends in a failed state, as we have proven. 
The latter means that no infinite computation is possible with the recursive rule, independent of the program in which it appears.
Otherwise, every non-terminating computation of the recursive rule in any program contains the maximally vicious computation, as we have proven, too.

The resulting devil's rules are often simple, 
e.g. non-recursive for single-recursion, %!!! 
and thus can be more easily inspected and analysed than their recursive counterparts.
Also, the maximally vicious computation can be performed as an instructive help for the programmer during debugging, since it exhibits the essence of non-termination.

The devil's advocate approach can be characterized as follows: 
It is concerned with universal (non-)termination, 
while most other work deals with termination in the given context of a specific program. In the latter, it is important to find out which non-terminating computations are unreachable, and this is a necessary complication that comes at a considerable cost.
It leads to combinatorial explosion and requires guessing of suitable abstractions.
The search for feasible execution paths that is typical for most research on non-termination 
is a dynamic analysis technique, 
while the construction of the devil's rule is straightforward and a static analysis technique. 

We have also introduced preliminary sufficient conditions for termination and non-termination that are directly derived from the devil's rules. At the moment, this compares favorable with other approaches, where the search for suitable invariants and recurrences involves indeterminism, heuristics and approximation techniques. It should be noted that the conditions are quite similar. Currently it is not clear if the simplicity is due to universal termination that we are interested in or if it will vanish once our conditions become more tight. 

We think the main appeal of our approach lies in the particulary simple construction of the malicious program, providing a finite witness for (non-)termination. Indeed, as we have shown, the malicious program can form an alternative basis for dynamic and static termination analysis.

Last but not least, our approach works well in a concurrent distributed language setting. Universal termination is an important issue there, since a malicious program produced by a devil's advocate could be introduced into the distributed environment to cause harm. A concrete example would be denial-of-service attacks.

\noindent {\bf Future Work}. We consider this paper as a starting point, many directions for future work are possible. First of all, we clearly should extend the applicability of the devil's rule construction to recursive simpagation and propagation rules as well as to mutual recursion. Secondly, we would like to improve the results on static analysis with additional conditions for termination and non-termination. We suspect there is a close relationship with existing approaches concerning that aspects of our work.
We also think that the presentation of the proofs could be made more accessible if an operational semantics more adequate for this kind of analysis can be found. 
Recent CHR semantics such as \cite{betz_raiser_fru_execution_model_iclp10} could provide a starting point.

In the context of CHR, ranking functions have been used to prove termination and complexity bounds
for bounded goals \cite{fru_complexity_kr02,fru_complexity2_entcs02}. 
Do devil's advocate rules respect such rankings? Can we derive boundedness conditions from them?
A classic analysis result for CHR is a decidable, sufficient and necessary condition for confluence of terminating programs \cite{abd_fru_completion_cp98}. There are also conditions for confluence of non-terminating programs. 
Do devil's rules respect confluence? 
We also think that the restriction to confluent programs may provide for additional conditions concerning static analysis of (non-)termination. 

Our devil's advocate method should be applied to other programming languages and paradigms. Logic programming languages such as Prolog and concurrent constraint languages should be especially suitable, since they are predecessors of CHR.
In Prolog there exist successful tools for termination analysis, which could provide an environment for fruitful comparisons with our approach.

For non-declarative languages we are confident that the advantages of our new technique carry over to this setting, where loops dominate over recursion as a language construct. A concrete starting point for this line of investigation might be to explore the relationship with the work \cite{chen2014proving}, as mentioned in the introduction.

We can regard the devil's rules as a finite, concise and compact representation of queries of finite and infinite size. It still could be worthwhile to derive (some of) these queries as counter-examples from the malicious program. This could be useful for the user, but also foster comparison with other approaches in the field. Similarily, the given program context could be taken into account to see if infinite computations according to the devil's rule are possible at all.

In the end, our devil's advocate method, once fully understood and explored, might work best when combined with existing approaches, hopefully combining their advantages and leveling out their disadvantages.

\bibliographystyle{abbrv}
\bibliography{devils,tfall2005,biblio}

\end{document}